\documentclass[12pt]{article}
\usepackage{amsmath}
\usepackage{bm}

\begin{document}
\begin{center}
\begin{large}
{\bf Perturbation of the $ns$ energy levels of the hydrogen atom in rotationally invariant noncommutative space}
\end{large}
\end{center}

\centerline { Kh. P. Gnatenko \footnote{E-Mail address: khrystyna.gnatenko@gmail.com}, Yu. S. Krynytskyi  \footnote{E-Mail address: yurikryn@gmail.com}, V. M. Tkachuk \footnote{E-Mail address: voltkachuk@gmail.com} }
\medskip
\centerline {\small \it Ivan Franko National University of Lviv, Department for Theoretical Physics,}
\centerline {\small \it 12 Drahomanov St., Lviv, 79005, Ukraine}

\abstract{Noncommutative space which is rotationally invariant is considered. The hydrogen atom is studied in this space. We exactly find the leading term in the asymptotic expansion of the corrections to the $ns$ energy levels over the small parameter of noncommutativity.


  PACS number(s): 11.90.+t, 11.10.Nx, 03.65.-w.}
\section{Introduction}

Noncommutativity has recently received a considerable interest owing to development of String Theory and Quantum Gravity (see, for instance, \cite{Witten,Doplicher}). It is worth noting that the idea that space might have a noncommutative structure has a long history. This idea was suggested by Heisenberg and later formalized by Snyder \cite{Snyder}.

Canonical version of noncommutative space is characterized by the following commutation relations for the coordinate and momentum operators
 \begin{eqnarray}
[X_{i},X_{j}]=i\hbar\theta_{ij},\label{form101}\\{}
[X_{i},P_{j}]=i\hbar\delta_{ij},\\{}
[P_{i},P_{j}]=0,\label{form10001}{}
\end{eqnarray}
where $\theta_{ij}$ is a constant antisymmetric matrix. Many physical
problems have been studied in this space (see, for instance, \cite{Gnatenko1} and references therein). Among these problems the hydrogen atom has been considered
\cite{Chaichian,Ho,Chaichian1,Chair,Stern,Zaim2,Adorno,Khodja}. In
\cite{Chaichian} the authors found the corrections to the energy levels of hydrogen atom
up to the first order in the parameter of
noncommutativity. In that article the
corrections to the Lamb shift within the noncommutative quantum
electrodynamics theory were also obtained. In \cite{Ho} the hydrogen atom was studied as a two-particle system. The authors considered the case when the particles of
opposite charges feel opposite noncommutativity. In \cite{Chair} the quadratic
Stark effect was examined. Shifts
in the spectrum of hydrogen atom in noncommutative space were presented in \cite{Stern}. In \cite{Zaim2} the hydrogen atom energy levels were calculated in the framework of the
noncommutative Klein-Gordon equation. The
Dirac equation with a Coulomb field was studied in
noncommutative space in \cite{Adorno,Khodja}.

 Hydrogen atom was also considered in the case of
space-time noncommutativity
 \cite{Balachandran,Stern1,Moumni1,Moumni,Zaim}, phase-space noncommutativity \cite{Djemai,Kang,Alavi,Bertolami}, $\kappa$-space-time (see, for example, \cite{Harikumar}).

It is important to note that there is a problem of rotational symmetry breaking in a canonical version of noncommutative space (see, for instance, \cite{Chaichian,Balachandran1}). In order to solve this problem different classes of noncommutative algebras were considered (see, for instance, \cite{Gnatenko} and references therein).

In the previous work \cite{Gnatenko} in order to preserve the rotational symmetry in noncommutative space we considered the idea to generalize the constant antisymmetric matrix $\theta_{ij}$ to a tensor. We proposed to construct this tensor with the help of additional coordinates and defined it as follows
 \begin{eqnarray}
\theta_{ij}=\frac{\alpha}{\hbar}(a_{i}b_{j}-a_{j}b_{i}),\label{form130}
\end{eqnarray}
where $\alpha$ is a dimensionless constant, and $a_{i}$, $b_{i}$ are governed by the harmonic
oscillators
 \begin{eqnarray}
 H_{osc}=\frac{(p^{a})^{2}}{2m}+\frac{(p^{b})^{2}}{2m}+\frac{m\omega^{2} a^{2}}{2}+\frac{m\omega^{2}b^{2}}{2}.\label{form104}
 \end{eqnarray}
The parameter of noncommutativity is thought to be of the order of the Planck scale. Therefore, we put \begin{eqnarray}
\sqrt{\frac{\hbar}{m\omega}}=l_{p},
 \end{eqnarray}
where $l_{p}$ is the Planck length. It was also proposed to consider the limit $\omega\rightarrow\infty$. In this case harmonic oscillator put into the ground state remains in it.

In this article according to the previous suggestion presented in \cite{Gnatenko} we consider the following commutation relations
\begin{eqnarray}
[X_{i},X_{j}]=i\alpha(a_{i}b_{j}-a_{j}b_{i}),\label{form131}\\{}
[X_{i},P_{j}]=i\hbar\delta_{ij},\\{}
[P_{i},P_{j}]=0.\label{form13331}{}
\end{eqnarray}
The coordinates $a_{i}$, $b_{i}$, and momenta $p^{a}_{i}$, $p^{b}_{i}$ satisfy the ordinary commutation relations $[a_{i},a_{j}]=0$, $[a_{i},p^{a}_{j}]=i\hbar\delta_{ij}$, $[b_{i},b_{j}]=0$, $[b_{i},p^{b}_{j}]=i\hbar\delta_{ij}$, also $[a_{i},b_{j}]=[a_{i},p^{b}_{j}]=[b_{i},p^{a}_{j}]=[p^{a}_{i},p^{b}_{j}]=0$. It is worth noting that $a_{i}$, $b_{i}$ commute with $X_{i}$ and $P_{i}$ and therefore $\theta_{ij}$ given by (\ref{form130}) commutes with $X_{i}$ and $P_{i}$ too. So, $X_{i}$, $P_{i}$ and $\theta_{ij}$ satisfy the same commutation relations as in the case of the canonical version of noncommutativity. Besides, algebra (\ref{form131})-(\ref{form13331}) is manifestly rotationally invariant \cite{Gnatenko}.

Also in \cite{Gnatenko} the energy levels of the hydrogen atom were studied in rotationally invariant noncommutative space (\ref{form131})-(\ref{form13331}). In article \cite{Gnatenko} we obtained an approximate result for the corrections to these levels in noncommutative space. In the present article we find exactly the leading term in the asymptotic expansion of the correction to the $ns$ energy levels over the small parameter of noncommutativity.

The article is organized as follows. In Section \ref{rozd2} the corrections to the $ns$ energy levels are considered. The leading term in the asymptotic expansion of the corrections to the $ns$ energy levels over the small parameter of noncommutativity is calculated in Section \ref{rozd3}. Conclusions are presented in Section \ref{rozd4}.

\section{Corrections to the $ns$ Energy Levels of the Hydrogen Atom }\label{rozd2}

Let us consider the perturbation of the $ns$ energy levels of the hydrogen atom in rotationally invariant noncommutative space.

The Hamiltonian of the hydrogen atom reads
 \begin{eqnarray}
 H_{h}=\frac{P^{2}}{2M}-\frac{e^{2}}{R},
 \label{form888}
  \end{eqnarray}
where the coordinates $X_{i}$ and momenta $P_{i}$  satisfy (\ref{form131})-(\ref{form13331}) and $R=\sqrt{\sum_{i}X_{i}^{2}}$.
Besides, defining the tensor of noncommutativity as (\ref{form130}), we have to take into account
additional terms that correspond to the harmonic oscillator
(\ref{form104}) and consider the total Hamiltonian as follows
 \begin{eqnarray}
H=H_{h}+H_{osc}.\label{form13}
 \end{eqnarray}

Using representation
\begin{eqnarray}
X_{i}=x_{i}+\frac{1}{2}[{\bm{\theta}}\times{\bf p}]_{i},\label{form10}\\
P_{i}=p_{i},
\end{eqnarray}
where the coordinates $x_{i}$ and momenta $p_{i}$ satisfy the ordinary commutation relations $[x_{i},x_{j}]=[p_{i},p_{j}]=0$,
$[x_{i},p_{j}]=i\hbar\delta_{ij}$, and
 \begin{eqnarray}
{\bm{\theta}}=\frac{\alpha}{\hbar} [{\bf a}\times{\bf b}],\label{form9088}
\end{eqnarray}
we can rewrite Hamiltonian (\ref{form13}) in the following form
\begin{eqnarray}
 H=H_{0}+V,\label{form600}
\end{eqnarray}
where  $H_{0}=H_{h}^{(0)}+H_{osc}$. Here $H_{h}^{(0)}=\frac{p^{2}}{2M}-\frac{e^{2}}{r}$ is the Hamiltonian of the hydrogen atom in the ordinary commutative space and $V$ is the perturbation caused by the noncommutativity of coordinates
\begin{eqnarray}
V=-\frac{e^{2}}{\sqrt{r^{2}-({\bm{\theta}}\cdot{\bf L})+\frac{1}{4}[{\bm{\theta}}\times{\bf p}]^{2}}}+\frac{e^{2}}{r},\label{form600456}
\end{eqnarray}
with $r=\sqrt{\sum_{i}x_{i}^{2}}$.

 It is worth noting that in \cite{Gnatenko}, expanding the perturbation $V$ caused by the noncommutativity of coordinates (\ref{form600456}) over ${\bm \theta}$ and using the perturbation theory, we have faced the problem of divergence of the corrections to the $ns$ energy levels.  Therefore in this article
we propose the way to  find exactly expression for the leading term in the asymptotic expansion of the corrections to the $ns$ energy levels over the small parameter of noncommutativity.

As was mention in the previous section, in the limit $\omega\rightarrow\infty$  harmonic oscillator is always in the ground state. Therefore, let us find the corrections to the $ns$ energy levels in the case when the harmonic oscillator is in the ground state. According to the perturbation theory we have
\begin{eqnarray}
\Delta E_{ns}=\left\langle\psi^{(0)}_{n,0,0,\{0\},\{0\}}({\bf r},{\bf a},{\bf b})\left|\frac{e^{2}}{r}-\frac{e^{2}}{\sqrt{r^{2}-({\bm{\theta}}\cdot{\bf L})+\frac{1}{4}[{\bm{\theta}}\times{\bf p}]^{2}}}\right|\psi^{(0)}_{n,0,0,\{0\},\{0\}}({\bf r},{\bf a},{\bf b})\right\rangle,\nonumber\\
{}\label{form721}
 \end{eqnarray}
where $\psi^{(0)}_{n,0,0,\{0\},\{0\}}({\bf r},{\bf a},{\bf b})$ are the eigenfunctions of the unperturbed Hamiltonian $H_{0}$.
Note that $H_{h}^{(0)}$ commutes with $H_{osc}$. Therefore, the eigenfunctions $\psi^{(0)}_{n,0,0,\{0\},\{0\}}({\bf r},{\bf a},{\bf b})$ read
\begin{eqnarray}
\psi^{(0)}_{n,0,0,\{0\},\{0\}}({\bf r},{\bf a},{\bf b})=\psi_{n,0,0}({\bf r})\psi^{a}_{0,0,0}({\bf a})\psi^{b}_{0,0,0}({\bf b}), \label{form321}
 \end{eqnarray}
 where $\psi_{n,0,0}({\bf r})$ are well known eigenfunctions of the hydrogen atom in ordinary space,  $\psi^{a}_{0,0,0}({\bf a})$, $\psi^{b}_{0,0,0}({\bf b})$ are the eigenfunctions of the three-dimensional harmonic oscillators in the ground state. Note that all these eigenfunctions are real, therefore further we omit notation of complex conjugation of this functions in integrals.

  It can be shown that $({\bm{\theta}}\cdot{\bf L})$ commutes with $[{\bm{\theta}}\times{\bf p}]^{2}$ and $r^{2}$. Also, it is worth noting that $({\bm{\theta}}\cdot{\bf L})\psi^{(0)}_{n,0,0,\{0\},\{0\}}({\bf r},{\bf a},{\bf b})=0$. Therefore, we can write
\begin{eqnarray}
\Delta E_{ns}=\left\langle\psi^{(0)}_{n,0,0,\{0\},\{0\}}({\bf r},{\bf a},{\bf b})\left|\frac{e^{2}}{r}-\frac{e^{2}}{\sqrt{r^{2}+\frac{1}{4}[{\bm{\theta}}\times{\bf p}]^{2}}}\right|\psi^{(0)}_{n,0,0,\{0\},\{0\}}({\bf r},{\bf a},{\bf b})\right\rangle.\nonumber\\
{}\label{form90333}
 \end{eqnarray}

For convenience we introduce dimensionless coordinates ${\bf a}^\prime={\bf a}/{l_{p}}$, ${\bf b}^\prime={\bf b}/{l_{p}}$, where $l_{p}$ is the Planck length. Therefore, ${\bm{\theta}}$ reads \begin{eqnarray}
{\bm{\theta}}=\frac{\alpha l_{p}^{2}}{\hbar}{\bm{\theta}}^\prime,\label{form9033}
\end{eqnarray}
where
 \begin{eqnarray}
 {\bm{\theta}}^\prime=[{\bf a}^\prime\times{\bf b}^\prime].
 \end{eqnarray}
Also, we use the following notation ${\bf r}^\prime=\sqrt{\frac{2}{\alpha}}\frac{{\bf r}}{l_{p}}$. As a consequence we can rewrite $\Delta E_{ns}$ as follows
\begin{eqnarray}
\Delta E_{ns}=\frac{\chi^{2}e^{2}}{a_{B}}I_{ns}(\chi),\label{form903}
 \end{eqnarray}
where
 \begin{eqnarray}
I_{ns}(\chi)=\int d{\bf a}^\prime \tilde{\psi}^{a}_{0,0,0}({\bf a}^\prime)\int d{\bf b}^\prime \tilde{\psi}^{b}_{0,0,0}({\bf b}^\prime)\int d{\bf r}^\prime\tilde{\psi}_{n,0,0}(\chi{\bf r}^\prime) \left(\frac{1}{r^\prime}\right.-\nonumber\\-\left.\frac{1}{\sqrt{(r^\prime)^{2}+[\bm{\theta}^\prime\times{\bf p}^\prime]^{2}}}\right)\tilde{\psi}_{n,0,0}(\chi{\bf r}^\prime)\tilde{\psi}^{a}_{0,0,0}({\bf a}^\prime)\tilde{\psi}^{b}_{0,0,0}({\bf b}^\prime),\label{form947}
 \end{eqnarray}
 here
 \begin{eqnarray}
 \chi=\sqrt{\frac{\alpha}{2}}\frac{l_p}{a_B},\label{form8407}
 \end{eqnarray}
 and $\tilde{\psi}^{a}_{0,0,0}({\bf a}^\prime)=\pi^{-\frac{3}{4}}e^{-\frac{(a^\prime)^2}{2}}$, $\tilde{\psi}^{b}_{0,0,0}({\bf b}^\prime)=\pi^{-\frac{3}{4}}e^{-\frac{(b^\prime)^2}{2}}$ are the dimensionless eigenfunctions corresponding to the harmonic oscillators, $\tilde{\psi}_{n,0,0}(\chi{\bf r}^\prime)=\sqrt{\frac{1}{\pi n^{5}}}e^{-\frac{\chi r^\prime}{n}}L_{n-1}^{1}\left(\frac{2\chi r^\prime}{n}\right) $ are the dimensionless eigenfunctions of the hydrogen atom, $L_{n-1}^{1}\left(\frac{2\chi r^\prime}{n}\right)$ are the generalized Laguerre polynomials.

Note that in the case of $\chi=0$ integral (\ref{form947}) has a finite value. Therefore, the asymptotic of $\Delta E_{ns}$ for $\chi\rightarrow0$ ($\alpha\rightarrow0$) is as follows
 \begin{eqnarray}
\Delta E_{ns}=\frac{\chi^{2}e^{2}}{a_{B}}I_{ns}(0).\label{form930}
 \end{eqnarray}

So, in order to find the asymptotic of $\Delta E_{ns}$ we have to find $I_{ns}(0)$.
It is worth mentioning that in the previous article \cite{Gnatenko} the corrections to the $ns$ energy levels were found up to linear fluctuations of $[\bm{\theta}^\prime\times{\bf p}^\prime]^{2}$. In the integral $I_{ns}(0)$ we replaced $\langle f(A)\rangle$ by $f(\langle A\rangle)$. Namely, the following replacement was considered
 \begin{eqnarray}
\left\langle\frac{1}{\sqrt{(r^\prime)^{2}+[\bm{\theta}^\prime\times{\bf p}^\prime]^{2}}}\right\rangle_{{\bf a}^{\prime}, {\bf b}^{\prime}}\longrightarrow\frac{1}{\sqrt{(r^\prime)^{2}+\left\langle[\bm{\theta}^\prime\times{\bf p}^\prime]^{2}\right\rangle_{{\bf a}^{\prime},{\bf b}^{\prime}}}},
 \end{eqnarray}
 where $\langle...\rangle_{{\bf a}^{\prime},{\bf b}^{\prime}}$ denotes $\langle\tilde{\psi}^{a}_{0,0,0}({\bf a}^{\prime})\tilde{\psi}^{b}_{0,0,0}({\bf b}^{\prime})|...|\tilde{\psi}^{a}_{0,0,0}({\bf a}^{\prime})\tilde{\psi}^{b}_{0,0,0}({\bf b}^{\prime})\rangle_{{\bf a}^{\prime},{\bf b}^{\prime}}$. Therefore an approximate result for corrections to the $ns$ energy levels was obtained.

 In this article we calculate the integral $I_{ns}(0)$ exactly.
First let us consider the integral over ${\bf r}^\prime$, namely
\begin{eqnarray}
I_{ns}(\chi,\bm{\theta}^\prime)=\int d{\bf r}^\prime\tilde{\psi}_{n,0,0}(\chi{\bf r}^\prime) \left(\frac{1}{r^\prime}-\frac{1}{\sqrt{(r^\prime)^{2}+[\bm{\theta}^\prime\times{\bf p}^\prime]^{2}}}\right)\tilde{\psi}_{n,0,0}(\chi{\bf r}^\prime).\label{form901}
\end{eqnarray}

It is convenient to use the momentum representation. We have
\begin{eqnarray}
I_{ns}(\chi,\bm{\theta}^\prime)=\nonumber\\=
\frac{1}{\chi^6}\int d{\bf p}^{\prime}\tilde{\psi}_{n,0,0}\left(\frac{{\bf p}^{\prime}}{\chi}\right)\left(\frac{1}{\sqrt {-\nabla_{p^{\prime}}^2}}-\frac{1}{\sqrt{-\nabla_{p^{\prime}}^2+[\bm{\theta}^{\prime}\times{\bf p}^{\prime}]^{2}}}\right)\tilde{\psi}_{n,0,0}\left(\frac{{\bf p}^{\prime}}{\chi}\right),\label{form932}
\end{eqnarray}
where $\nabla_{p^{\prime}}^2=\sum_{i}\frac{\partial^{2}}{(\partial p^{\prime}_i)^{2}}$. It is worth noting that the integral $I_{ns}(\chi,\bm{\theta}^\prime)$ does not depend on the direction of $\bm{\theta}^{\prime}$.  Therefore, we can write
\begin{eqnarray}
I_{ns}(\chi,\bm{\theta}^\prime)=I_{ns}(\chi,\theta^\prime)=\nonumber\\=
\frac{1}{4\pi\chi^6}\int d\Omega\int d{\bf p}^{\prime}\tilde{\psi}_{n,0,0}\left(\frac{{\bf p}^{\prime}}{\chi}\right)\left(\frac{1}{\sqrt {-\nabla_{p^{\prime}}^2}}-\frac{1}{\sqrt{-\nabla_{p^{\prime}}^2+[\bm{\theta}^{\prime}\times{\bf p}^{\prime}]^{2}}}\right)\tilde{\psi}_{n,0,0}\left(\frac{{\bf p}^{\prime}}{\chi}\right) =\nonumber\\=
\frac{1}{4\pi\chi^6}\int d\Omega\int d{\bf p}^{\prime}\tilde{\psi}_{n,0,0}\left(\frac{{\bf p}^{\prime}}{\chi}\right)\left(\frac{1}{\sqrt {-\nabla_{p^{\prime}}^2}}-\frac{1}{\sqrt{-\nabla_{p^{\prime}}^2+(\theta^{\prime})^{2}(p^{\prime})^{2}\sin^2\Theta}}\right)\tilde{\psi}_{n,0,0}\left(\frac{{\bf p}^{\prime}}{\chi}\right),\nonumber\\
{}\label{form933}
\end{eqnarray}
where $\theta^\prime=|\bm{\theta}^\prime|$, and $d\Omega=\sin\Theta d\Theta d\Phi$, $\Theta$ is an angle between vectors $\bm{\theta}^\prime$ and ${\bf p^\prime}$. Using the following substitution $\tilde{{\bf p}}=\kappa{\bf p}^{\prime}$, with $\kappa=\sqrt{\theta^{\prime}\sin\Theta}$, and returning to the coordinate representation, we can rewrite (\ref{form933}) in the following form

\begin{eqnarray}
I_{ns}(\chi,\theta^\prime)=\frac{\theta^{\prime}}{2}\int_{0}^{\pi} d\Theta \sin^2\Theta \int d\tilde{{\bf r}}\tilde{\psi}_{n,0,0}(\kappa\chi\tilde{{\bf r}})\left(\frac{1}{\tilde{r}}-\frac{1}{\sqrt{\tilde{r}^{2}+\tilde{p}^{2}}}\right)\tilde{\psi}_{n,0,0}(\kappa\chi{\bf \tilde{r}})=\nonumber\\=
\frac{\theta^{\prime}}{2}\int_{0}^{\pi} d\Theta \sin^2\Theta \int_{0}^{\infty} d \tilde{r}\tilde{r}^{2} \tilde{R}_{n,0}(\kappa\chi \tilde{r})\left(\frac{1}{\tilde{r}}-\frac{1}{\sqrt{\tilde{r}^{2}+p^{2}_{\tilde{r}}}}\right)\tilde{R}_{n,0}(\kappa\chi \tilde{r}),\nonumber\\
{}\label{form943}
\end{eqnarray}
here $\tilde{R}_{n,0}(\kappa\chi \tilde{r})=\sqrt{\frac{4}{n^{5}}}e^{-\frac{\kappa\chi \tilde{r}}{n }}L_{n-1}^{1}\left(\frac{2\kappa\chi \tilde{r}}{n}\right)$ is the dimensionless radial wave function of the hydrogen atom,  $p_{\tilde{r}}=-i\frac{1}{\tilde{r}}\frac{\partial}{\partial \tilde{r}}\tilde{r}$.

It is convenient to use the following notation
\begin{eqnarray}
S_{ns}(\kappa\chi)=\nonumber\\=4\int_{0}^{\infty} d \tilde{r}\tilde{r}^{2} e^{-\frac{\kappa\chi \tilde{r}}{n }}L_{n-1}^{1}\left(\frac{2\kappa\chi \tilde{r}}{n}\right)\left(\frac{1}{\tilde{r}}-\frac{1}{\sqrt{\tilde{r}^{2}+p^{2}_{\tilde{r}}}}\right)e^{-\frac{\kappa\chi \tilde{r}}{n }}L_{n-1}^{1}\left(\frac{2\kappa\chi \tilde{r}}{n}\right),\label{form902}
\end{eqnarray}
 and rewrite $I_{ns}(\chi,\theta^\prime)$ as follows
\begin{eqnarray}
I_{ns}(\chi,\theta^\prime)=
\frac{\theta^{\prime}}{2n^{5}}\int_{0}^{\pi} d\Theta \sin^2\Theta S_{ns}(\kappa\chi).\label{form56}
\end{eqnarray}

Now, returning to (\ref{form930}), and taking into account that
\begin{eqnarray}
I_{ns}(0)=\langle I_{ns}(0,\theta^\prime)\rangle_{{\bf a}^{\prime},{\bf b}^{\prime}},\\
I_{ns}(0,\theta^\prime)=
\frac{\theta^{\prime}}{2n^{5}}\int_{0}^{\pi} d\Theta \sin^2\Theta S_{ns}(0)=\frac{\pi\theta^\prime}{4n^{5}}S_{ns}(0),\label{form956}
\end{eqnarray}
we obtain
\begin{eqnarray}
\Delta E_{ns}=\frac{\pi\langle\theta^{\prime}\rangle\chi^{2}e^{2}}{4a_{B}n^{5}}S_{ns}(0),\label{form91212}
\end{eqnarray}
where
\begin{eqnarray}
\langle\theta^{\prime}\rangle=\langle\tilde{\psi}^{a}_{0,0,0}({\bf a}^{\prime})\tilde{\psi}^{b}_{0,0,0}({\bf b}^{\prime})|\sqrt{\sum_{i}(\theta^{\prime}_{i})^{2}}|\tilde{\psi}^{a}_{0,0,0}({\bf a}^{\prime})\tilde{\psi}^{b}_{0,0,0}({\bf b}^{\prime})\rangle=1.\label{form20}
\end{eqnarray}
 Note that the result for $\langle\theta^{\prime}\rangle$ is expectable because of the way of introducing the dimensionless coordinates ${\bf a}^{\prime}$, ${\bf b}^{\prime}$.

It can be shown that
\begin{eqnarray}
S_{ns}(0)=S_{1s}(0)n^{2}.\label{form951}
\end{eqnarray}

Finally, using (\ref{form20}) and (\ref{form951}), we exactly obtain the following expression for the leading term in the asymptotic expansion of the corrections to the $ns$ energy levels
\begin{eqnarray}
\Delta E_{ns}=\frac{\pi\chi^{2}e^{2}}{4a_{B}n^{3}}S_{1s}(0).\label{form987}
\end{eqnarray}
So, in order to find (\ref{form987}) we have to calculate $S_{1s}(0)$.

\section{Calculation of the Leading Term in the Asymptotic Expansion of the Corrections the $ns$ Energy Levels}\label{rozd3}

As was shown in the previous Section, to find the leading term in the asymptotic expansion of the corrections the $ns$ energy levels we have to calculate $S_{1s}(0)$
\begin{eqnarray}
S_{1s}(0)=4\int_{0}^{\infty} d \tilde{r}\tilde{r}^{2} \left(\frac{1}{\tilde{r}}-\frac{1}{\sqrt{\tilde{r}^{2}+p_{\tilde{r}}^{2}}}\right).\label{form904}
\end{eqnarray}
For this purpose let us expand $1$ over the eigenfunctions of $\tilde{r}^{2}+p_{\tilde{r}}^{2}$
 \begin{eqnarray}
 1=\sum_{k=0}^{\infty}C_{k}\phi_{k},
  \label{form720}
  \end{eqnarray}
where $\phi_{k}$ are the eigenfunctions of $\tilde{r}^{2}+p_{\tilde{r}}^{2}$
 \begin{eqnarray}
 \phi_{k}=\sqrt{\frac{2k!}{\Gamma(k+\frac{3}{2})}}e^{-\frac{\tilde{r}^{2}}{2}}L^{\frac{1}{2}}_{k}(\tilde{r}^{2}),\label{form740}
  \end{eqnarray}
see, for instance, \cite{Yanez} and $C_{k}$ are the expansion coefficients
   \begin{eqnarray}
   C_{k}=\sqrt{\frac{2 k!}{\Gamma(k+\frac{3}{2})}}\int_{0}^{\infty}d\tilde{r}\tilde{r}^{2}e^{-\frac{\tilde{r}^{2}}{2}}L^{\frac{1}{2}}_{k}\left(\tilde{r}^{2}\right)=(-1)^{k}\sqrt{\frac{4\Gamma(k+\frac{3}{2})}{k!}}.
  \end{eqnarray}

As a result, the second term in (\ref{form904}) reads
 \begin{eqnarray}
 \int_{0}^{\infty}d\tilde{r}\tilde{r}^{2}\frac{1}{\sqrt{\tilde{r}^{2}+p_{\tilde{r}}^{2}}}= \sum_{k=0}^{\infty}\frac{C_{k}^{2}}{\sqrt{\lambda_{k}}},
 \label{form711}
 \end{eqnarray}
 where  $\lambda_{k}$ are the eigenvalues of $\tilde{r}^{2}+p_{\tilde{r}}^{2}$, namely $\lambda_{k}=2\left(2k+\frac{3}{2}\right)$.

The first term in (\ref{form904}) can be presented in the following form
 \begin{eqnarray}
\int_{0}^{\infty}d\tilde{r}\tilde{r}=\sum_{k=0}^{\infty}C_{k}I_{k},
 \label{form710}
 \end{eqnarray}
 where
 \begin{eqnarray}
 I_{k}=\sqrt{\frac{2k!}{\Gamma(k+\frac{3}{2})}}\int_{0}^{\infty}d\tilde{r} \tilde{r} e^{-\frac{\tilde{r}^{2}}{2}}L^{\frac{1}{2}}_{k}\left(\tilde{r}^{2}\right)=\nonumber\\=(-1)^{k}\sqrt{\frac{8 k!}{\pi\Gamma(k+\frac{3}{2})}}{}_{2}F_{1}\left(-k,\frac{1}{2};\frac{3}{2};2\right),
\end{eqnarray}
 here ${}_{2}F_{1}\left(-k,\frac{1}{2};\frac{3}{2};2\right)$ is the hypergeometric function. So, taking into account (\ref{form711}) and (\ref{form710}), we have
  \begin{eqnarray}
S_{1s}(0)=4\sum_{k=0}^{\infty}\left(C_{k}I_{k}-\frac{C_{k}^{2}}{\sqrt{\lambda_{k}}}\right)=\nonumber\\=
16\sqrt{\frac{2}{\pi}}\sum_{k=0}^{\infty}\frac{\Gamma(k+\frac{3}{2})}{k!}\left({}_{2}F_{1}\left(-k,\frac{1}{2};\frac{3}{2};2\right)-\sqrt{\frac{\pi}{8k+6}}\right). \label{form908}
\end{eqnarray}
Note that the two sums in $S_{1s}(0)$, namely
 \begin{eqnarray}
16\sqrt{\frac{2}{\pi}} \sum_{k=0}^{\infty}\frac{\Gamma(k+\frac{3}{2})}{k!}{}_{2}F_{1}\left(-k,\frac{1}{2};\frac{3}{2};2\right), \label{form909}\\
16\sqrt{\frac{2}{\pi}} \sum_{k=0}^{\infty}\frac{\Gamma(k+\frac{3}{2})}{k!}\sqrt{\frac{\pi}{8k+6}}, \label{form910}
  \end{eqnarray}
  are divergent. Nevertheless, the value of $S_{1s}(0)$ is finite. In order to work with the sums  (\ref{form909}) and (\ref{form910}) separately let us use an additional multiplier $\eta^{k}$ ($\eta<1$)
 \begin{eqnarray}
16\sqrt{\frac{2}{\pi}} \sum_{k=0}^{\infty}\frac{\Gamma(k+\frac{3}{2})}{k!}{}_{2}F_{1}\left(-k,\frac{1}{2};\frac{3}{2};2\right)\eta^{k}, \label{form911}\\
 16\sqrt{\frac{2}{\pi}}\sum_{k=0}^{\infty}\frac{\Gamma(k+\frac{3}{2})}{k!}\sqrt{\frac{\pi}{8k+6}}\eta^{k}. \label{form912}
  \end{eqnarray}
Note, that in the case of $\eta=1$ we obtain (\ref{form909}), (\ref{form910}).

First we consider sum (\ref{form912}). It is clear that
\begin{eqnarray}
\sqrt{\frac{\pi}{k+\frac{3}{4}}}=2\int_{0}^{\infty}dze^{-(k+\frac{3}{4})z^{2}}.\label{form7040}
\end{eqnarray}
Also, it can be shown that
\begin{eqnarray}
\sum_{k=0}^{\infty}\frac{\Gamma(k+\frac{3}{2})}{k!}t^{k}=\frac{\sqrt{\pi}}{2(1-t)^{\frac{3}{2}}}.\label{form7011}
\end{eqnarray}
As a result, taking into account (\ref{form7040}) and (\ref{form7011}), we obtain
\begin{eqnarray}
16\sqrt{2}\sum_{k=0}^{\infty}\frac{\Gamma(k+\frac{3}{2})}{k!\sqrt{8k+6}}\eta^{k}=
16\sum_{k=0}^{\infty}\frac{\Gamma(k+\frac{3}{2})}{k!\sqrt{\pi}}\eta^{k}\int_{0}^{\infty}dze^{-(k+\frac{3}{4})z^{2}}=\nonumber\\=
8\int_{0}^{\infty}dz\frac{e^{-\frac{3}{4}z^{2}}}{(1-\eta e^{-^{z^{2}}})^{\frac{3}{2}}}.\label{form7017}
\end{eqnarray}

Now let us consider sum (\ref{form911}). The hypergeometric function  $_{2}F_{1}(-k,\frac{1}{2};\frac{3}{2};2)$ can be presented in the following form
\begin{eqnarray}
_{2}F_{1}\left(-k,\frac{1}{2};\frac{3}{2};2\right)=\sum_{q=0}^{k}\frac{(-1)^{q}C_{k}^{q}2^{q}}{2q+1},\label{form7013}
\end{eqnarray}
where $C_{k}^{q}$ are the binomial coefficients.
It is clear that
\begin{eqnarray}
\frac{1}{2q+1}=\int_{0}^{1}dzz^{2q}.\label{form7014}
\end{eqnarray}
Therefore, using (\ref{form7013}) and (\ref{form7014}), we obtain
\begin{eqnarray}
{}_{2}F_{1}\left(-k,\frac{1}{2},\frac{3}{2},2\right)=\sum_{q=0}^{k}\int_{0}^{1}dz C_{k}^{q}(-2)^{q}z^{2q}=\int_{0}^{1}dz(1-2z^{2})^{k}.\label{form7012}
\end{eqnarray}

Finally, using (\ref{form7011}) and (\ref{form7012}), we can rewrite (\ref{form911}) as follows
\begin{eqnarray}
16\sqrt{\frac{2}{\pi}}\sum_{k=0}^{\infty}\frac{\Gamma(k+\frac{3}{2})}{k!}{}_{2}F_{1}\left(-k,\frac{1}{2},\frac{3}{2},2\right)\eta^{k}=\nonumber\\=
16\sqrt{\frac{2}{\pi}}\sum_{k=0}^{\infty}\frac{\Gamma(k+\frac{3}{2})}{k!}\eta^{k}\int_{0}^{1}dz(1-2z^{2})^{k}=8\sqrt{2}\int_{0}^{1}\frac{dz}{(1-\eta(1-2z^{2}))^{\frac{3}{2}}}.\label{form7015}
\end{eqnarray}
Splitting the integral (\ref{form7015}) into two integrals we have
\begin{eqnarray}
\int_{0}^{1}\frac{dz}{(1-\eta(1-2z^{2}))^{\frac{3}{2}}}=I_{1}(\eta)+I_{2}(\eta),
\end{eqnarray}
where
\begin{eqnarray}
I_{1}(\eta)=\int_{0}^{\frac{1}{\sqrt{2}}}\frac{dz}{(1-\eta(1-2z^{2}))^{\frac{3}{2}}},\label{form7090}\\
I_{2}(\eta)=\int_{\frac{1}{\sqrt{2}}}^{1}\frac{dz}{(1-\eta(1-2z^{2}))^{\frac{3}{2}}}.\label{form7016}
\end{eqnarray}
Note that the integral $I_{2}(\eta)$ has a finite value even for $\eta=1$. Putting $\eta=1$ in (\ref{form7016}), we obtain
\begin{eqnarray}
I_{2}(1)=\frac{\sqrt{2}}{8}.\label{form913}
\end{eqnarray}

Next let us rewrite (\ref{form7090}) in the form which is close to (\ref{form7017}). Using substitution $e^{-t^{2}}=1-2z^{2}$, we have
\begin{eqnarray}
I_{1}(\eta)=\frac{\sqrt{2}}{2}\int_{0}^{\infty}dt\frac{te^{-t^{2}}}{(1-e^{-t^{2}})^{\frac{1}{2}}(1-\eta e^{-t^{2}})^{\frac{3}{2}}}.\label{form7018}
\end{eqnarray}
As a result, taking into account  (\ref{form7017}), (\ref{form7015}), (\ref{form913}), (\ref{form7018}), we find
 \begin{eqnarray}
16\sqrt{\frac{2}{\pi}} \sum_{k=0}^{\infty}\frac{\Gamma(k+\frac{3}{2})}{k!}{}_{2}F_{1}\left(-k,\frac{1}{2};\frac{3}{2};2\right)\eta^{k}-
 16\sqrt{\frac{2}{\pi}}\sum_{k=0}^{\infty}\frac{\Gamma(k+\frac{3}{2})}{k!}\sqrt{\frac{\pi}{8k+6}}\eta^{k}=\nonumber\\=8\sqrt{2}I_{2}(\eta)+8\int_{0}^{\infty}dt\frac{te^{-t^{2}}-e^{-\frac{3}{4}t^{2}}(1- e^{-t^{2}})^{\frac{1}{2}}}{(1-e^{-t^{2}})^{\frac{1}{2}}(1-\eta e^{-t^{2}})^{\frac{3}{2}}}.\nonumber\\
 {}\label{form7019}
  \end{eqnarray}
 Note that the integral in (\ref{form7019}) has a finite value even for $\eta=1$ and this integral can be easy calculated. Consequently, putting $\eta=1$ in (\ref{form7019}), and taking into account (\ref{form908}), (\ref{form913}), we find
\begin{eqnarray}
S_{1s}(0)=2+8\int_{0}^{\infty}dt\frac{te^{-t^{2}}-e^{-\frac{3}{4}t^{2}}\sqrt{1- e^{-t^{2}}}}{(1-e^{-t^{2}})^{2}}=1.72006\ldots\label{form7020}
\end{eqnarray}

Now, using (\ref{form987}), we obtain
\begin{eqnarray}
\Delta E_{ns}\simeq1.72\frac{\pi\chi^2e^{2}}{4a_{B}n^{3}}.
\end{eqnarray}

Finally, taking into account  (\ref{form9033}) and (\ref{form8407}), we have the following corrections written in term of the parameter of noncommutativity
\begin{eqnarray}
\Delta E_{ns}\simeq1.72\frac{\hbar\langle\theta\rangle\pi e^{2}}{8a_{B}^{3}n^{3}},\label{form7999}
\end{eqnarray}
where
\begin{eqnarray}
\langle\theta\rangle=\langle\psi^{a}_{0,0,0}({\bf a})\psi^{b}_{0,0,0}({\bf b})|\sqrt{\sum_{i}\theta_{i}^{2}}|\psi^{a}_{0,0,0}({\bf a})\psi^{b}_{0,0,0}({\bf b})\rangle=\frac{\alpha l_{p}^{2}}{\hbar},
\end{eqnarray}
and $\bm{\theta}$ is given by (\ref{form9088}).

At the end of this section we would like to note that result (\ref{form7999}) and approximate result calculated in \cite{Gnatenko} differ in multiplier $\frac{\pi}{4}$. Therefore, an upper bound for the parameter of noncommutativity obtained in \cite{Gnatenko} will not be significantly changed. Nevertheless, it is worth noting that in this article we have found exact result for the expression for the leading term in the asymptotic expansion of the corrections to the $ns$ energy levels over the small parameter of noncommutativity (\ref{form987}), where $S_{1s}(0)$ is given by (\ref{form7020}).

 \section{Conclusion}\label{rozd4}

In this article we have considered the idea to construct rotationally invariant noncommutative algebra by the generalization of a constant antisymmetric matrix to a tensor defined by additional coordinates (\ref{form130}). In this rotationally invariant noncommutative space the hydrogen atom has been studied.

The main result of this paper is exactly found expression for the leading term in the asymptotic expansion of the corrections to the $ns$ energy levels over the small parameter of noncommutativity (\ref{form987}), where $S_{1s}(0)$ is given by (\ref{form7020}).
It is worth noting that dependence of this term on the parameter of noncommutativity is proportional to $\langle\theta\rangle$, whereas the corrections to the energy levels with $l>1$ are proportional to $\langle\theta^{2}\rangle$ \cite{Gnatenko}. Therefore, we can conclude that $ns$ energy levels are more sensitive to the noncommutativity of coordinates.

\section{Acknowledgements}
The authors thank Dr. A. A. Rovenchak for a careful reading of the manuscript.

\end{document}